\documentclass[sigconf,letterpaper]{acmart}

\usepackage{xspace}
\newcommand{\BfPara}[1]{\vspace{1mm}{\noindent\bf #1.}\xspace}
\usepackage{tikz}
\newcommand*\cced[1]{\tikz[baseline=(char.base)]{
         \node[shape=circle,fill,inner sep=0.8pt] (char) {\textcolor{white}{#1}};}}

\usepackage{booktabs}

\usepackage{amsmath}
\usepackage{graphicx}
\usepackage{makecell}
\usepackage{array}
\newcolumntype{L}[1]{>{\raggedright\let\newline\\\arraybackslash\hspace{0pt}}m{#1}}
\newcolumntype{C}[1]{>{\centering\let\newline\\\arraybackslash\hspace{0pt}}m{#1}}
\newcolumntype{R}[1]{>{\raggedleft\let\newline\\\arraybackslash\hspace{0pt}}m{#1}}

\setcopyright{none}
\renewcommand\footnotetextcopyrightpermission[1]{} 
\begin{document}

\title[Do Content Management Systems Impact the Security of Free Content Websites?]{Do Content Management Systems Impact the  Security of Free Content Websites? A Correlation Analysis}

\author{Mohammed Alaqdhi}
\affiliation{%
   \institution{University of Central Florida}
   \city{Orlando}
   \state{FL}
   \country{USA}}
\email{malqadhi@Knights.ucf.edu}

\author{Abdulrahman	Alabduljabbar}
\affiliation{%
   \institution{University of Central Florida}
   \city{Orlando}
   \state{FL}
   \country{USA}}
\email{jabbar@Knights.ucf.edu}

\author{Kyle	Thomas}
\affiliation{%
   \institution{University of Central Florida}
   \city{Orlando}
   \state{FL}
   \country{USA}}
\email{kthomas4031@Knights.ucf.edu}

\author{Saeed	Salem}
\affiliation{%
   \institution{Qatar University}
   \city{Doha}
   \country{Qatar}}
\email{saeed.salem@qu.edu.qa}

\author{DaeHun Nyang}
\affiliation{%
   \institution{Ewha Womans University}
   \city{Seoul}
   \country{Republic of Korea}}
\email{saeed.salem@ndsu.edu}

\author{David Mohaisen}
\affiliation{%
   \institution{University of Central Florida}
   \city{Orlando}
   \state{FL}
   \country{USA}}
\email{mohaisen@ucf.edu}

\begin{abstract}
This paper investigates the potential causes of the vulnerabilities of free content websites to address risks and maliciousness. Assembling more than 1,500 websites with free and premium content, we identify their content management system (CMS) and malicious attributes. We use frequency analysis at both the aggregate and per category of content (books, games, movies, music, and software), utilizing the unpatched vulnerabilities, total vulnerabilities, malicious count, and percentiles to uncover trends and affinities of usage and maliciousness of CMS{'s} and their contribution to those websites. Moreover, we find that, despite the significant number of custom code websites, the use of CMS{'s} is pervasive, with varying trends across types and categories. Finally, we find that even a small number of unpatched vulnerabilities in popular CMS{'s} could be a potential cause for significant maliciousness. 


\end{abstract}

\maketitle

\section{Introduction}

Today, free content websites are an essential part of the Internet, providing ample resources to users in the form of free books, movies, software, and games, among others. Free content websites have always been a focal point of debate and covered in various studies~\cite{akhawe2010towards,garfinkel2002web,calzavara2016content}. The main questions around the study of free content websites have been their security and privacy: what are the direct and indirect costs associated with using those websites? Those costs have been studied by contrasting free content websites with premium websites--websites that provide similar content but charge fees--across multiple analysis dimensions, including their vulnerabilities in the code base, infrastructure utilization, and the richness of their privacy policies~\cite{adepoju2020human,pan2016cspautogen,verkijika2018quality}. 

For instance, in some of the prior work, it was reported that there is a higher level of maliciousness in free content websites than in premium websites~\cite{mohaisen2015amal,kosba2014adam}, as reported in various scanners (e.g., \url{virustotal.com}), which makes the free content websites unsafe to visit by their users~\cite{alabduljabbar2022poster}. Digital certificates, a key component in ensuring the confidentiality and integrity of the communication between browsers and those websites~\cite{alrawi2016chains}, are shown to be problematic in many ways. For example, those websites tend to have mismatched domain names as a result of poor website migration or even are expired~\cite{alabduljabbar2022understanding}. The privacy policies of those websites are also shown to be limited or may not necessarily cover various essential policy elements that are expected in general and are shown in the privacy policies of premium content websites~\cite{alabduljabbar2022measuring}. 

Despite the importance and coverage of the various studies in the literature~\cite{alabduljabbar2022measuring,alabduljabbar2022understanding}, they fall short in various aspects, particularly in understanding and identifying the root cause of the lack of security and privacy in free content websites. The contrast provided in the literature highlights that free content websites are a source of lurking risks and vulnerabilities that could expose users and their data to significant security costs. Whether it is in the eventual detection as malicious, the lack of expressiveness in their privacy policy, or the significantly lax certificate qualities, all in contrast to the premium content websites. However, there is a lack of a study that looks into various potential contributors to the vulnerability, particularly in those websites codes, to better understand a mitigation strategy for the associated risks. 

To address this gap in the existing literature on the understanding of the security of free content websites, we revisit the security analysis of free content websites through code-based analysis. The critical insight we utilize for our study is that the security of any website is best understood by understanding the codebase of its content, and the shared code in particular. In essence, we hypothesize that many of the vulnerabilities associated with those websites could be caused by a repeated pattern in their codebase due to the utilization of third-party components, libraries, or just insecure coding practices, as is the case with many web technologies. We find that we can understand the repeated patterns by studying the utilization of third-party content management systems (CMS{'s}), which are heavily utilized in today's website development. 

\BfPara{Contributions} In this paper, we contribute to the state-of-the-art by analyzing and contrasting the security of free content websites through the lenses of CMS analysis using 1,562 websites. We annotate the websites with their malicious attributes and systematically evaluate the role of CMS as a contributing factor. We find that a significant number of the websites ($\approx$44\%) use CMS{'s}, which comes with vulnerabilities and contributes to maliciousness. We find that the use pattern of CMS{'s} is unique across different types of websites and categories. The top-used CMS{'s} have several aspects in common, such as unpatched vulnerabilities, which help explain the maliciousness of websites using them.


\BfPara{Organization} The rest of this paper is as follows. In \autoref{sec:related}, we review the related work. In \autoref{sec:data}, we review our dataset and its annotation. In \autoref{sec:methods}, we provide an overview of the methods utilized in this paper. In \autoref{sec:results}, we provide the results and the discussion. Finally, in \autoref{sec:final}, we provide the conclusion and recommendation for future research or work.

\section{Related Work}\label{sec:related}

In the following, we sample and review the most related pieces of prior work to the work presented in this study.

\BfPara{Online Website Analysis} Researchers have held that diverse constituents might be subject to increased risks when using free content websites, given the evolution of online services and web applications. These risks have been examined across various website features, including digital certificates, content, and addressing infrastructure.~\cite{alabduljabbar2022understanding}. In another study, component and website-level analyses were conducted to understand vulnerabilities utilizing two main off-the-shelf tools, VirusTotal and Sucuri~\cite{alabduljabbar2022poster}, linking free content websites to significant threats.

\BfPara{Privacy Practices Reporting} Mindful of the implicit security cost, another work has looked into the interplay between privacy policies and the quality of those websites. Namely, the prior work examined user comprehension of risks linked to service use through privacy policy understanding~\cite{alabduljabbar2022measuring}. The researchers passed several filtered privacy policies into a custom pipeline that annotates the policies against various categories (e.g., first and third-party usage, data retention)~\cite{jayanthi2011xgraphticsclus}. The authors found that the privacy policies of free content websites are vague, lack essential policy elements, or are lax in specifying the responsibilities of the service provider (website owner) against possible compromise and exposure of user data. On the other hand, they found that the privacy policies of the premium content websites are more transparent and elaborate about reporting their practices on data gathering, sharing, and retention~\cite{alabduljabbar2022measuring}.

\BfPara{Tracking and Website Structure} 
Another study has contributed to this field by revealing the tracking mechanisms of corporate ownership~\cite{libert2015exposing}. To comprehend the web tracking phenomenon and subsequently craft material policies to regulate it, the authors argued that it is imperative to know the actual degree and reach of corporations that may be subject to the increased regulations. The most significant finding in this research was that 78.07 percent of websites within Alexa's top million instigated third-party HTTP requests to the domain owned by Google. Furthermore, the researchers observed that the overall trend shown by past surveys is not only that many of the users of websites value privacy but also that the present privacy state online denotes an area of material anxiety. Concerning measurement, the same study highlights that the level of tracking on the web is on the rise and does not show indications of abating.

\section{Dataset and Data Annotation}\label{sec:data}

\BfPara{Websites} For this study, we compiled a dataset that contains 1,562 websites, with 834 free content websites and 728 premium websites, which have been used in prior work~\cite{alabduljabbar2022poster,alabduljabbar2022understanding,alabduljabbar2022measuring}. In selecting those websites, we consider their popularity while maintaining a balance per the sub-category of a website. To determine the popularity of a website, we used the results of search engines Bingo, DuckDuckGo, and Google as a proxy, where highly ranked websites are considered popular. To balance the dataset, we undertook a manual verification approach to vet each website across the sub-category (see below). Namely, we sorted the websites into five categories based on the content they predominantly serve: software, music, movies, games, or books. The following are the free and premium content websites count per category: books (154 free, 195 premium), games (80 free, 113 premium), movies (331 free, 152 premium), music (83 free, 86 premium), and software (186 free, 182 premium). 

\BfPara{Dataset annotation} For our analysis, we augment the dataset in various ways. We primarily focused on information reflecting the exposure to the risk of users~\cite{alabduljabbar2022understanding}.   
We determine whether a website is malicious or benign using the VirusTotal API~\cite{VirusTotal}. VirusTotal is a framework that offers cyber threat detection, which helps us analyze, detect, and correlate threats while reducing the required effort through automation. Specifically, the API allowed us to identify malicious IP addresses, domains, or URLs associated with the websites we use for augmentation.


\BfPara{\em CMS{'s}} Since this work aims to understand the role of software (CMS, in particular) used across websites and its contribution to threat exposure, we follow a two-step approach: (1) website crawling and (2) manual inspection and annotation. First, we crawl each of the websites and inspect its elements to find the source folder for the website. From the source folder, we list the source and content for each website to identify the CMS used to develop this website. This approach requires us to build a database of the different available CMS{'s} to allow automation of the annotation through regular expression matching. We cross-validate our annotation utilizing existing online tools used for CMS detection. We use CMS-detector~\cite{CMSDetect} and w3techs~\cite{W3Techs}, two popular tools, to extract the CMS{'s} used for the list of websites. For automation, we build a wrapper that prepares the query with the website, retrieves the response of the CMS used from the corresponding tool, and compares it to the manually identified set in the previous step. Among the CMS{'s} identified, WordPress is the most popular, followed by Drupal, Django, Next.js, Laravel, CodeIgniter, and DataLife. In total, we find 77 unique CMS{'s} used across the different websites, not including websites that rely on a custom-coded CMS.

\BfPara{\em Vulnerabilities} Our dataset's final augmentation and annotation are the vulnerability count and patching patterns. For each CMS, we crawl the results available in various portals concerning the current version of the CMS to identify the associated vulnerability. Namely, we crawl such information from cvedetails~\cite{CVEDetails}, snyk.io~\cite{snyk}, openbugbounty~\cite{openbugbounty}, and Wordfence~\cite{Wordfence}. Finally, to determine whether a vulnerability is patched or not (thus counting the number of unpatched vulnerabilities), we query cybersecurity-help~\cite{cybersecurity-help}.

\section{Analysis Methods}\label{sec:methods}



The key motivation behind our analysis is to understand the potential contribution of CMS{'s} to the (in)security of free content websites, which has been established already in the prior work, as highlighted in section~\ref{sec:related}. To achieve this goal, we pursue two directions. The first is a holistic analysis geared toward understanding the distribution of various features associated with free content and premium websites (combined). The second is a fine-grained analysis that considers the per-category analysis of vulnerabilities. In essence, our study utilizes frequency analysis of various features to understand trends and affinities and a holistic view of vulnerabilities. The features we utilize are as follows:

\begin{enumerate}\itemsep=1mm
    \item {\bf CMS} This feature signifies the industry name of the content management system utilized by the free content website, premium content website, or both. 
    \item {\bf Count} This feature signifies the total number of websites that utilize the given CMS for their operation. In particular, we assume each given website utilizes only one CMS, which has been the case in our analysis. \item {\bf Percent} This feature signifies the normalization of the count feature by the total number of studied websites. We use the percentage to understand a relative order of the CMS{'s} contribution that is easier to interpret.
    \item {\bf Malicious count (MC)} This feature is calculated per CMS. It highlights the total number of websites utilizing the given CMS deemed malicious. For our maliciousness check, we utilize the output of VirusTotal, where a website is deemed malicious if at least one scanner has flagged it as malicious. 
    \item {\bf Malicious percentage per CMS (MPCP)} This feature signifies the normalized MC by the count feature. It highlights the significance (as a percentage) of the malicious websites for the given CMS. It highlights the actual relative contribution of the CMS to the maliciousness of websites taking into account their relative representation in our dataset. 
    \item {\bf Malicious percentage (MP)} This feature signifies the MC feature normalized by the total MC value (i.e., the overall number of malicious websites) by capturing a given CMS{'s} relative contribution to the total number of malicious websites. It signifies the contribution irrespective of the representation of that CMS in our dataset. The gap between MPCP and MP signifies whether a given CMS is more secure in the abstract or not. 
    \item {\bf Total vulnerabilities (TV)} This feature signifies the total number of vulnerabilities associated with the given CMS. 
    \item {\bf Unpatched vulnerabilities (UV)} This feature signifies the total number of unpatched vulnerabilities associated with the given CMS.
    \item {\bf Correlation analysis} This feature identifies the relationship between the CMS{'s} and the maliciousness of the sites. For that, we use the Pearson correlation coefficient, defined as $\rho_{X,Y}=\frac{\text{cov}(X,Y)}{\sigma_X\sigma_Y}$. Here, $X$ is a random variable associated with free content/premium content type (malicious vs. benign), and $Y$ is a random variable capturing the CMS{'s} associated with the given type. 
\end{enumerate}

\section{Results and Discussion} \label{sec:results}
In this section, we highlight the results of this study, including an overall analysis overview, category-based analysis results, and discussion, and putting it together through correlation analysis. 


\subsection{Overall Analysis Results and Discussion}\label{sec:holistic}

First, we explore the distribution of the various features outlined earlier in a holistic manner, considering the free vs. premium labels of the websites. The results are shown in Table~\ref{tab:holistic}, and we make the following observations. 

\begin{description}
    \item[The Prevalence of Malicious Websites.] The total number of malicious websites is 525 out of 1,561 websites, corresponding to 33.63\% of them. This number is surprisingly large, especially in contrast to general website maliciousness levels, which are estimated at 1\%\footnote{\url{https://ophtek.com/what-are-malicious-websites/}}. 

\item[Various Vulnerability Types in Popular CMS's.] In terms of vulnerability, the maliciousness of those websites corresponds to 2,760 vulnerabilities in the CMS{'s} those websites employ. Of those vulnerabilities, 145 are unpatched at the time of our scanning. While small as a percentage (only 5.25\%), we note that some of those unpatched vulnerabilities are associated with the most popular CMS{'s} our dataset. For example, one unpatched vulnerability is associated with WordPress, which is used by more than 24\% of websites and is associated with 32\% of the total number of malicious websites. This supports our hypothesis on the role of CMS{'s} as an amplification avenue of vulnerabilities and associated impact, where a single vulnerability could be utilized to contaminate a large number of websites and recruit them into malicious endeavors. 

\item[The Prevalence of Custom Codes.] We observe that a majority of the websites (883, or 56.6\%) in our dataset use custom code, with 30.5\% (or 269) of them being malicious. Custom-coded websites made up 51.2\% of the total malicious websites. In contrast, while the websites that used CMS{'s} represented 43.4\% of all websites, they had 48.8\% of all malicious websites, which corresponds to 37.8\% maliciousness among those that utilize CMS{'s}.

\item[While Large, Our Estimates are Conservative.] Our estimate of the role of CMS{'s} serves only a lower bound, as we do not consider the potential for shared codes among custom websites (i.e., websites that do not use a standard CMS). Those websites might be reusing cross-website codes, which could amplify the vulnerabilities. 

\item[Vulnerabilities Vary by Website Type] We observe a range of percentages of vulnerabilities and maliciousness across the website groups utilizing different CMS{'s}, where that percentage sometimes exceeds 40\% (well above the average) even with major CMS{'s}; e.g., WordPress (44.3\%), Next.js (53.85\%), and  Shopify (70\%). These results show a significant trend in the maliciousness of the websites based on their platforms.
\end{description}

\subsection{Category-based Analysis}
The main results provided in section~\ref{sec:holistic} are profound, although they do not look into the individual categories and how they differ (if at all). To help answer this question, we conduct the same analysis we had in section~\ref{sec:holistic}, but per category; for books, games, movies, music, and software. Our analysis provides a contrast against the mean (\textsection\ref{sec:holistic}) and group (free vs. premium).

\begin{table}[t]
\centering
\def\arraystretch{0.95}
\caption{Distribution of the combined free and premium websites across different CMS{'s}. Studied distribution characteristics are for each CMS: the percentage among all websites (percent), the count, malicious count (MC), malicious per CMS websites count (MPCP), the malicious percentage among the total websites (MP), the total number of identified vulnerabilities with the given CMS (TV), and the total number of unpatched vulnerabilities for the given CMS (UV).}\label{tab:holistic}
\scalebox{0.8}{
\begin{tabular}{L{0.14\textwidth}R{0.04\textwidth}R{0.04\textwidth}R{0.04\textwidth}R{0.05\textwidth}R{0.04\textwidth}R{0.04\textwidth}R{0.04\textwidth}}
\Xhline{1\arrayrulewidth}
CMS & \# & \% & MC & MPCP & MP & TV & UV \\
\Xhline{1\arrayrulewidth}
Custom code      & 883 & 56.57 & 269 & 30.46 & 17.23 & -- & -- \\
WordPress        & 379 & 24.28 & 168 & 44.33 & 10.76 & 8 & 1 \\
Zendesk          & 26 & 1.67 & 11 & 42.31 & 0.70 & 2 & 2 \\
Drupal           & 25 & 1.60 & 3 & 12.00 & 0.19 & 228 & 0 \\
Adobe EM         & 22 & 1.41 & 1 & 4.55 & 0.06 & 93 & 0 \\
Shopify          & 20 & 1.28 & 14 & 70.00 & 0.90 & 0 & 0 \\
Magento          & 18 & 1.15 & 5 & 27.78 & 0.32 & 210 & 3 \\
Next.js          & 13 & 0.83 & 7 & 53.85 & 0.45 & 9 & 0 \\
Laravel          & 9 & 0.58 & 2 & 22.22 & 0.13 & 9 & 1 \\
vBulletin        & 9 & 0.58 & 3 & 33.33 & 0.19 & 0 & 0 \\
HubSpot CMS      & 8 & 0.51 & 5 & 62.50 & 0.32 & 3 & 0 \\
Bigcommerce      & 6 & 0.38 & 0 & 0.00 & 0.00 & 20 & 0 \\
Django Framework & 6 & 0.38 & 1 & 16.67 & 0.06 & 1 & 0 \\
Salesforce C360  & 6 & 0.38 & 0 & 0.00 & 0.00 & 1 & 0 \\
Gatsby           & 5 & 0.32 & 2 & 40.00 & 0.13 & 1 & 0 \\
IPS Community    & 5 & 0.32 & 2 & 40.00 & 0.13 & 3 & 0 \\
Joomla           & 5 & 0.32 & 1 & 20.00 & 0.06 & 83 & 2 \\
Oracle CX        & 5 & 0.32 & 0 & 0.00 & 0.00 & 25 & 0 \\
Salesforce Cloud & 5 & 0.32 & 5 & 100 & 0.32 & 56 & 0 \\
Sitecore CMS     & 5 & 0.32 & 2 & 40.00 & 0.13 & 19 & 1 \\
Others          & 101 & 6.47 & 24 & 23.76 & 1.54 & 1,989 & 135 \\
\Xhline{1\arrayrulewidth}
Total           & 1,561 & 100 & 525 & -- & 33.63 & 2,760 & 145 \\
\Xhline{1\arrayrulewidth}
\end{tabular}}
\end{table}

\begin{table}[h!]
\centering
\def\arraystretch{0.95}
\caption{Distribution of free vs. premium {\bf books content websites} across different CMS{'s}. Studied distribution characteristics are for each CMS; Keys are as in \autoref{tab:holistic}.}\label{tab:books}
\begin{tabular}{L{0.14\textwidth}R{0.04\textwidth}R{0.04\textwidth}R{0.04\textwidth}R{0.05\textwidth}R{0.04\textwidth}}
\Xhline{1\arrayrulewidth}
\multicolumn{6}{c}{Free Content Websites}\\
\Xhline{1\arrayrulewidth}
CMS & \# & \% & MC & MPCP & MP    \\
\Xhline{1\arrayrulewidth}
Custom code      & 115 & 75.16 & 31 & 26.96 & 20.26 \\
WordPress        & 22 & 14.38 & 10 & 45.45 & 6.54 \\
Drupal           & 3 & 1.96 & 0 & 0.00 & 0.00 \\
Django Framework & 2  & 1.31  & 1 & 50.00 & 0.65 \\
vBulletin        & 2 & 1.31 & 1 & 50.00 & 0.65 \\
Others           & 9 & 5.88 & 3 & 33.33 & 1.96 \\
\Xhline{1\arrayrulewidth}
Total            & 153 & 100 & 46 & -- & 30.07 \\
\Xhline{1\arrayrulewidth}
\multicolumn{6}{c}{Premium Websites}\\
\Xhline{1\arrayrulewidth}
Custom code     & 84 & 43.08 & 19 & 22.62 & 9.74 \\
WordPress       & 46 & 23.59 & 12 & 26.09 & 6.15 \\
Shopify         & 10 & 5.13 & 7 & 70.00 & 3.59 \\
Drupal          & 7 & 3.59 & 0 & 0.00 & 0.00 \\
Magento         & 6 & 3.08 & 2 & 33.33 & 1.03 \\
Others          & 42 & 21.53 & 13 & 30.95 & 6.67 \\
\Xhline{1\arrayrulewidth}
Total           & 195 & 100 & 53 & -- & 27.18 \\
\Xhline{1\arrayrulewidth}
\end{tabular}
\end{table}

\BfPara{General Observations} Before delving into the specific analysis of each category, we make the following broad observations. {\bf (1)} We notice that while the use of CMS{'s} is common among both the free content and premium websites, the usage follows different patterns: whereby the number of CMS{'s} utilized by the free content websites is small, it is prominent in the case of premium websites, with a more significant heavy-tailed distribution (i.e., a significant number of the CMS{'s} have a minimal representation in terms of the websites that utilize them). This is very well-captured in the ``others'' row in every table, where we combine the CMS{'s} with 1-2 websites. We observe that ``others'' in the case of premium websites is significantly more than that in the free content websites part of the table. {\bf (2)} Across the different websites, we observe inconsistent patterns concerning the division between custom code and CMS: where it is significantly greater in the case of free content websites vs. premium for books (75\% vs. 43\%), movies (86\% vs. 51\%), music (65\% vs. 51\%), and movies (59\% vs. 30\%), the pattern does not hold for games (47\% vs. 56\%).

\BfPara{\cced{1} Books} The results in \autoref{tab:books} show that there are 153 free content websites and 195 premium websites. With 348 websites, 149 use a CMS, and 199 use custom code. Under this category, 46 (30.8\%) of free content and 53 (27.2\%) of premium websites are malicious. In total, 99 (28.45\%) of the book websites category are malicious. This result shows that slightly more free content websites are malicious. In contrast, both types of websites have a malicious percentage that is less than the average (33.63\%) per \autoref{tab:holistic}. Interestingly, the free content websites have a 39.5\% chance of being malicious vs. a 30.6\% chance for the premium.  

Does the ranking of the CMS{'s} persist in both the free content and the premium websites? While the top CMS is the same in both cases, others in the top 4 for the free content (ordered) are Drupal, Django, vBulletin vs. Shopify, Drupal, and Magento for premium. Shopify is the most malicious CMS (percentage-wise) with 70\%. It is used only in the premium books category, in contrast to the top (count-wise) malicious CMS (WordPress) used in both.

\begin{table}[h!]
\centering\def\arraystretch{0.95}
\caption{Distribution of free vs. premium {\bf games content websites} across different CMS{'s}. Studied distribution characteristics are for each CMS; Keys are as in \autoref{tab:holistic}.}\label{tab:game}
\begin{tabular}{L{0.14\textwidth}R{0.04\textwidth}R{0.04\textwidth}R{0.04\textwidth}R{0.05\textwidth}R{0.04\textwidth}}
\Xhline{1\arrayrulewidth}
\multicolumn{6}{c}{Free Content Websites}\\
\Xhline{1\arrayrulewidth}
CMS & \# & \% & MC & MPCP & MP    \\
\Xhline{1\arrayrulewidth}
Custom code     & 38 & 47.50 & 22 & 57.89 & 27.50 \\
WordPress       & 34 & 13.75 & 18 & 52.94 & 22.50 \\
DataLife Engine & 2 & 2.50 & 2 & 100 & 2.50 \\
vBulletin       & 2 & 2.50 & 0 & 0.00 & 0.00 \\
Discuz!         & 1 & 1.25 & 0 & 0.00 & 0.00 \\
Others          & 3 & 3.75 & 1 & 33.33 & 1.25 \\
\Xhline{1\arrayrulewidth}
Total           & 80 & 100 & 43 & -- & 53.75   \\
\Xhline{1\arrayrulewidth}
\multicolumn{6}{c}{Premium Websites}\\
\Xhline{1\arrayrulewidth}
Custom code & 64 & 56.64 & 15 & 23.44 & 13.27 \\
WordPress   & 22 & 19.47 & 8 & 36.36 & 7.08 \\
Magento     & 4 & 3.54 & 2 & 50.00 & 1.77 \\
Zendesk     & 4 & 3.54 & 1 & 25.00 & 0.88 \\
Bigcommerce & 2 & 1.77 & 0 & 0.00 & 0.00 \\
Others & 17 & 15.04 & 9 & 52.94 & 7.96 \\
\Xhline{1\arrayrulewidth}
Total       & 113 & 100 & 35 & -- & 30.97  \\
\Xhline{1\arrayrulewidth}
\end{tabular}
\end{table}

\BfPara{\cced{2} Games} The results in \autoref{tab:game} show that there are 80 free content and 113 premium websites. With 193 websites, 91 use a CMS, and 102 use custom code. Under this category, the malicious count in free content is 43 (53.75\%) against 35 (31\%) in premium. The total number of malicious websites amounted to 78, or slightly more than 40\%. These results show that significantly more free content websites are malicious. In contrast, both types of websites have a malicious percentage that is close to or significantly higher than the average (33.63\%) per \autoref{tab:holistic}. Interestingly, for their representation in our data, the free websites have a 50\% chance of being malicious when using a CMS compared to about 40\% in the case of the premium websites.


While the top CMS is the same in both cases, others in the top 4 for free content (ordered) are DataLife Engine, vBulletin, and Discuz! vs. Magento, Zendesk, and Bigcommerce for premium games websites. DataLife Engine is the most malicious CMS (percentage-wise) at 100\%. It is used only in the free games category, in contrast to the top (count-wise) malicious CMS (WordPress) used in both categories.

\begin{table}[t]
\centering\def\arraystretch{0.95}
\caption{Distribution of free vs. premium {\bf movies content websites} across various CMS{'s}. Studied distribution characteristics are for each CMS; Keys are as in \autoref{tab:holistic}.}\label{tab:movie}
\begin{tabular}{L{0.14\textwidth}R{0.04\textwidth}R{0.04\textwidth}R{0.04\textwidth}R{0.05\textwidth}R{0.04\textwidth}}
\Xhline{1\arrayrulewidth}
\multicolumn{6}{c}{Free Content Websites}\\
\Xhline{1\arrayrulewidth}
CMS & \# & \% & MC & MPCP & MP    \\
\Xhline{1\arrayrulewidth}
Custom code      & 285 & 86.10 & 105 & 36.84 & 31.72 \\
Wordpress        & 34 & 10.27 & 17 & 50.00 & 5.14 \\
Django Framework & 2 & 0.60 & 0 & 0.00 & 0.00 \\
Laravel          & 2 & 0.60 & 1 & 50.00 & 0.30 \\
DataLife Engine  & 1 & 0.30 & 1 & 100 & 0.30 \\
Others           & 7 & 2.11 & 4 & 57.14 & 1.21 \\
\Xhline{1\arrayrulewidth}
Total            & 331 & 100 & 128 &--  & 38.67  \\
\Xhline{1\arrayrulewidth}
\multicolumn{6}{c}{Premium Websites}\\
\Xhline{1\arrayrulewidth}
Custom code & 78 & 51.32 & 6 & 7.69 & 3.95 \\
WordPress   & 18 & 11.84 & 5 & 27.78 & 3.29 \\
Zendesk     & 11 & 7.24 & 5 & 45.45 & 3.29 \\
Adobe EM    & 6 & 3.95 & 0 & 0.00 & 0.00 \\
Drupal      & 4 & 2.63 & 2 & 50.00 & 1.32 \\
Others      & 35 & 23.03 & 5 & 14.29 & 3.29 \\
\Xhline{1\arrayrulewidth}
Total       & 152 & 100 & 23 &  --& 15.13 \\
\Xhline{1\arrayrulewidth}
\end{tabular}
\end{table}

\BfPara{\cced{3} Movies} The results in \autoref{tab:movie} show that there are 331 free content websites and 152 premium content websites. With 483 websites, 120 use a CMS, and 363 use custom code. Under this category, there are 128 (38.67\%)malicious free content websites compared to 23 malicious premium websites (15.13\%). In total, 151 (30.26\%)  of the websites in the movies category are malicious. These results show that significantly more free content websites are malicious. In contrast, both types of websites have a malicious percentage that is slightly less than the average of (33.63\%) per \autoref{tab:holistic}. Interestingly, free content sites have close to a 40\% chance of being malicious vs. a 15.13\% chance for premium movie websites.

The top CMS is the same in both categories. Others in the top 4 for the free content (ordered) are Django Framework, Laravel, DataLife Engine vs. Zendesk, Adobe Experience Manager, and Drupal for premium. There is a draw between WordPress and Laravel for the most malicious (as a percentage) CMS used in free websites with 50\%. Drupal is the most malicious used only in premium movies with a percentage of 50\%. On the other hand, the top (count-wise) malicious CMS is (WordPress) used in free and premium movies.

\begin{table}[h!]
\centering\def\arraystretch{0.95}
\caption{Distribution of free vs. premium {\bf music content websites} across different CMS{'s}; Keys are as in \autoref{tab:holistic}.}\label{tab:music}
\begin{tabular}{L{0.14\textwidth}R{0.04\textwidth}R{0.04\textwidth}R{0.04\textwidth}R{0.05\textwidth}R{0.04\textwidth}}
\Xhline{1\arrayrulewidth}
\multicolumn{6}{c}{Free Content Websites}\\
\Xhline{1\arrayrulewidth}
CMS & \# & \% & MC & MPCP & MP    \\
\Xhline{1\arrayrulewidth}
Custom code & 54 & 65.06 & 24 & 44.44 & 28.92 \\
WordPress   & 18 & 21.69 & 5 & 27.78 & 6.02 \\
Drupal      & 2 & 2.41 & 0 & 0.00 & 0.00 \\
MediaWiki   & 2 & 2.41 & 0 & 0.00 & 0.00 \\
Shopify     & 2 & 2.41 & 1 & 50.00 & 1.20 \\
Others      & 5 & 6.02 & 2 & 40.00 & 2.41 \\
\Xhline{1\arrayrulewidth}
Total       & 83 & 100 & 32 &  --& 38.55  \\
\Xhline{1\arrayrulewidth}
\multicolumn{6}{c}{Premium Websites}\\
\Xhline{1\arrayrulewidth}
Custom code & 44 & 51.16 & 7 & 15.91 & 8.14 \\
WordPress   & 19 & 22.09 & 2 & 10.53 & 2.33 \\
Zendesk     & 4 & 4.65 & 2 & 50.00 & 2.33 \\
Gatsby      & 3 & 3.49 & 1 & 33.33 & 1.16 \\
Oracle CX   & 2 & 2.33 & 0 & 0.00 & 0.00 \\
Others      & 14 & 16.28 & 3 & 21.43 & 3.49 \\
\Xhline{1\arrayrulewidth}
Total       & 86 & 100  & 15 & -- & 17.44 \\
\Xhline{1\arrayrulewidth}
\end{tabular}
\end{table}

\BfPara{\cced{4} Music} The results in \autoref{tab:music} show that there are 83 free content websites and 86 premium content websites. With 169 websites, 98 use custom code, while 71 use a CMS. Under this category, 32 (38.6\%) of free content sites and 15 (17.44\%) of premium content sites are malicious. In total, 47 (27.8\%) of the music category websites are malicious. This result shows that significantly more free content websites are malicious. Both types of websites have a malicious percentage that is less than the average. For their representation in the dataset, the free websites have a nearly 40\% chance of being malicious vs. nearly 20\% chance in the premium websites.

The top CMS is (WordPress) in the free and premium content, followed by (ordered)  Drupal, MediaWiki, and Shopify in free content. Zendesk, Gatsby, and Oracle CX Commerce are the top in the premium websites category. We also note that the most malicious CMS (percentage-wise) is Shopify in free music and Zendesk in premium music with the same percentage of 50\%. While the top malicious CMS (count-wise) is (WordPress) and is used in both free and premium.


\begin{table}[t]
\centering\def\arraystretch{0.95}
\caption{Distribution of free vs. premium {\bf software content websites} across different CMS{'s}. Studied distribution characteristics are for each CMS; Keys are as in \autoref{tab:holistic}.}\label{tab:software}
\begin{tabular}{L{0.14\textwidth}R{0.04\textwidth}R{0.04\textwidth}R{0.04\textwidth}R{0.05\textwidth}R{0.04\textwidth}}
\Xhline{1\arrayrulewidth}
\multicolumn{6}{c}{Free Content Websites}\\
\Xhline{1\arrayrulewidth}
CMS & \# & \% & MC & MPCP & MP    \\
\Xhline{1\arrayrulewidth}
WordPress       & 111 & 59.68 & 81 & 72.97 & 43.55 \\
Custom code     & 69 & 37.10 & 33 & 47.83 & 17.74 \\
Contentteller   & 1 & 0.54 & 0 & 0.00 & 0.00 \\
IPS Community   & 1 & 0.54 & 1 & 100 & 0.54 \\
Jimdo           & 1 & 0.54 & 0 & 0.00 & 0.00 \\
Others          & 3 & 1.61 & 1 & 33.33 & 0.54 \\
\Xhline{1\arrayrulewidth}
Total           & 186 & 100  & 116 &  -- & 62.37 \\
\Xhline{1\arrayrulewidth}
\multicolumn{6}{c}{Premium Websites}\\
\Xhline{1\arrayrulewidth}
WordPress   & 55 & 30.22 & 10 & 18.18 & 5.49 \\
Custom code & 52 & 28.57 & 7 & 13.46 & 3.85 \\
Adobe EM    & 14 & 7.69 & 0 & 0.00 & 0.00 \\
Drupal      & 7 & 3.85 & 0 & 0.00 & 0.00 \\
Next.js     & 5 & 2.75 & 4 & 80.00 & 2.20 \\
Others      & 49 & 26.92 & 13 & 26.53 & 7.14 \\
\Xhline{1\arrayrulewidth}
Total       & 182 & 100 & 34 & -- & 18.68 \\
\Xhline{1\arrayrulewidth}
\end{tabular}
\end{table}

\begin{figure*}[htb]
    \centering
    \includegraphics[width=1\linewidth]{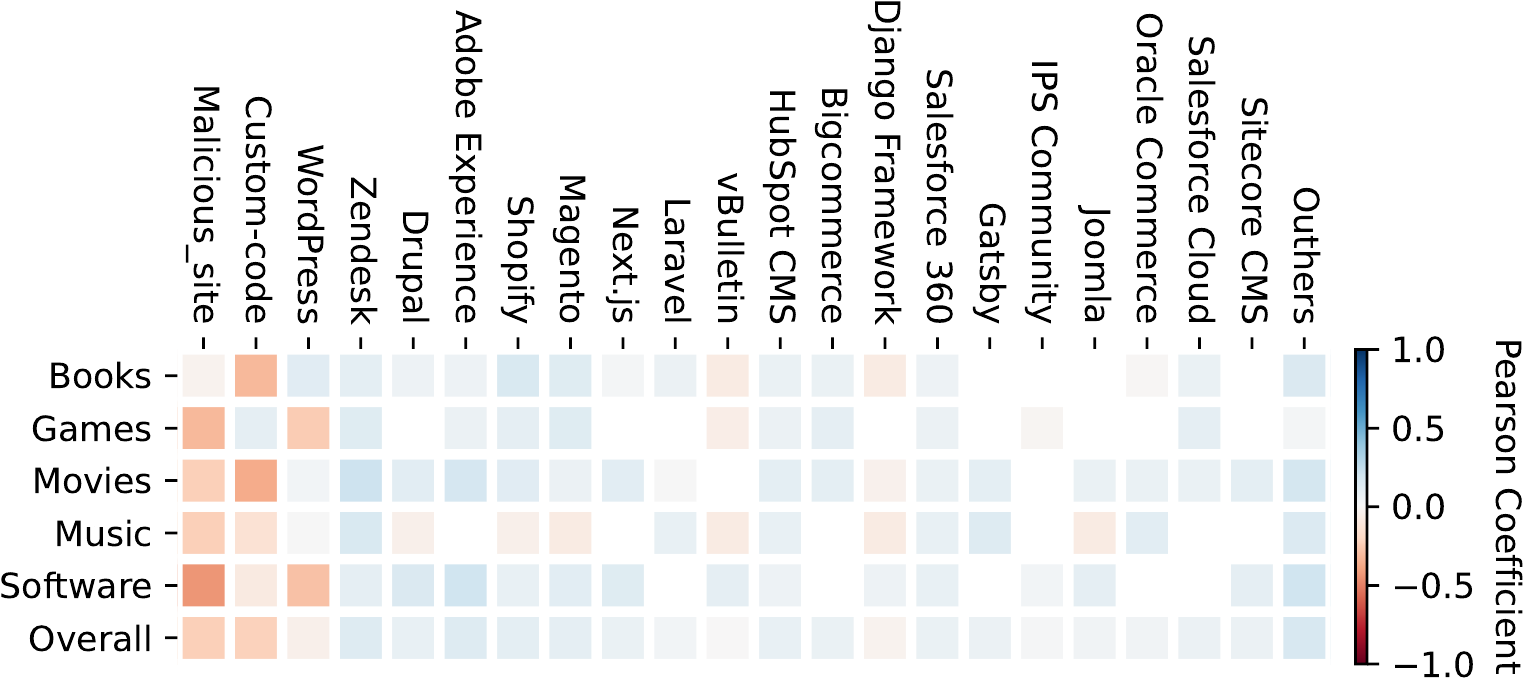}
    \caption{The Pearson correlation coefficient shows the linear relation between two features. Shown are the most contributing CMS{'s} based on Table 1 and the five content categories. The figure depicts the relation between the maliciousness of the websites and the different categories. The yellow color means that this feature strongly relates to free categories. In contrast, the blue color reflects a strong relationship with the premium content. The white color represents the weak relation.}
    \label{fig:heatmap}
\end{figure*}

\BfPara{\cced{5} Software} The results in \autoref{tab:software} show that there are 186 free content websites and 182 premium content websites. With 368 websites, 247 use a CMS, while 121 use custom code. Under this category, there are 116 (62.4\%) malicious free content websites, against 34 (18.7\%) malicious premium content websites. In total, 150 websites (40.76\%) of the software websites are malicious. These results show that significantly more free content websites are malicious. At the same time, both types have a malicious percentage that is more than the average of (33.63\%) per \autoref{tab:holistic}. Interestingly, the free websites have a more than 60\% chance of being malicious vs. near 20\% chance for the premium websites.

Unique to this category, we notice that the top code-base is not the custom code, but WordPress, which deviates from the last four categories. Moreover, the other top (ordered) CMS{'s} are Contentteller, IPS Community Suite, and Jimdo, among free content websites, vs. Adobe Experience Manager, Drupal, and Next.js for premium. Contentteller is shown to be the most malicious CMS (percentage-wise) with a 100\% chance and is used only in the free software category. In contrast to the top (count-wise) malicious CMS (WordPress) used in both free and premium.


\subsection{Putting it Together: Discussion}
\BfPara{\cced{1} Correlation Heatmap} The correlation heat map is shown in \autoref{fig:heatmap}. We find that most malicious sites are free content websites based on the correlation heat map. We also find that the free software websites are the most malicious, and this is shown in \autoref{tab:software} with (62.37\%) malicious percent. In contrast, the relation between the maliciousness of a website and the premium category is relatively weak. We also find that the premium category uses more CMS{'s} than the free websites. In the end, we find that free content websites using custom code are the most malicious. The second most are the sites using WordPress, which are likely to be malicious for both the free and premium categories. Premium websites using Zendesk and Shopify are the most malicious among the other premium websites.

\BfPara{\cced{2} Further Discussion}
Based on the previous results, we infer that CMS websites have a higher malicious percentage than custom code websites. \autoref{tab:holistic} shows that 30.46\% of the custom code websites are malicious against 38.55\% of the websites that use CMS{'s}. We also find that the free content sites have the highest vulnerabilities and maliciousness compared to premium websites in the per-category comparison. We also notice that websites that use specific CMS{'s}, such as WordPress, Shopify, Next.js, Gatsby, and Sitecore,  have a high chance of being malicious. It will be hard to generalize these results among the CMS{'s} that only occur once or twice, even if they have a 50\% chance of being malicious.

One outcome of this study is the assessment of the risk of CMS{'s} that are generally associated with malicious websites. Taking five as a threshold of occurrence with a malicious percent higher than 30\%, Shopify is considered high-risk, with 20 websites and 70\% of them being malicious. When applying the same criterion to other CMS{'s}, we find that Next.js (53.85\%, 13 times), WordPress (44.33\%, 379 times), and Zendesk (42.31\%, 26 times) are high-risk. We highlight the possibility of reducing the attack surface of websites by not using a high-risk CMS or by fixing the CMS to restrict these vulnerabilities. 

We noticed that the CMS with a lower malicious percentage has the highest number of total vulnerabilities but the lowest number of unpatched vulnerabilities. We highlight a pattern to use in practice: those with lower unpatched vulnerabilities are likely CMS{'s} that provide good maintenance and apply the latest security standards. One can recommend reducing the risk of websites using CMS{'s} using the same insights. It has been argued that this could be accomplished by having ongoing monitoring and management of the free content websites~\cite{ostroushko2015restricting}.

\section{Conclusion and Future Work} \label{sec:final}
Free content websites are an exciting web component. Our study shows various analyses to uncover specific risks associated with those websites in contrast to premium and highlights the significant challenges with free content websites in terms of increased vulnerabilities to maliciousness. Although well-established that free content websites are more likely to be malicious, we tie this likelihood to their utilization of CMS{'s}, in aggregate and at a per-category analysis. Recognizing this problem and the potential role CMS{'s} play in websites security, it is essential to generalize this insight to a larger number of websites, contrast those trends to other general websites (besides the free content vs. premium), and conduct measurements over time to capture the security dynamics. Moreover, it would be interesting to explore the utilization of CMS's in specific domains (e.g., healthcare~\cite{alkinoon2021measuring,alkinoon2021security}) and explore they contribute to their security. 

\subsection*{Acknowledgement} This work is supported in part by NRF grant 2016K1A1A2912757. 


\balance

\end{document}